\documentclass[conference]{IEEEtran}
\IEEEoverridecommandlockouts
\usepackage{cite}
\usepackage{amsmath,amssymb,amsfonts}
\usepackage{algorithmic}
\usepackage{graphicx}
\usepackage{textcomp}
\usepackage{xcolor}

\usepackage{amsmath}
\usepackage{amsthm}
\usepackage{booktabs}
\usepackage{algorithm}
\usepackage{algorithmic}
\usepackage{enumitem}

\usepackage{newfloat}
\usepackage{listings}
\usepackage{bibentry}

\def\BibTeX{{\rm B\kern-.05em{\sc i\kern-.025em b}\kern-.08em
    T\kern-.1667em\lower.7ex\hbox{E}\kern-.125emX}}
\begin{document}


\title{CineAGI: Character-Consistent Movie Creation through LLM-Orchestrated Multi-Modal Generation and Cross-Scene Integration}

\author{
\IEEEauthorblockN{Tianyidan Xie\textsuperscript{1}, Mingjie Wang\textsuperscript{2}, Qiang Tang\textsuperscript{3}, Feixuan Liu\textsuperscript{4}, Rui Ma\textsuperscript{5}, Lanjun Wang\textsuperscript{6}, Zili Yi\textsuperscript{1,*}\thanks{*Corresponding author: yi@nju.edu.cn}}
\\
\IEEEauthorblockA{\textsuperscript{1}Nanjing University, \textsuperscript{2}Zhejiang Sci-Tech University, \textsuperscript{3}University of British Columbia,\\
\textsuperscript{4}Beijing Shuzhimei Technology Co., Ltd, \textsuperscript{5}Jilin University, \textsuperscript{6}Tianjin University}
}

\maketitle

\begin{abstract}
Automated movie creation requires coordinating multiple characters, modalities, and narrative elements across extended sequences---a challenge that existing end-to-end approaches struggle to address effectively. We present \textbf{CineAGI}, a hierarchical movie generation framework that decomposes this complex task through specialized multi-agent orchestration. Our framework employs three key innovations: (1) a multi-agent narrative synthesis module where specialized LLM agents collaboratively generate comprehensive cinematic blueprints with character profiles, scene descriptions, and cross-modal specifications; (2) a decoupled character-centric pipeline that maintains identity consistency through instance-level tracking and integration while enabling flexible multi-character composition; and (3) a hierarchical audio-visual synchronization mechanism ensuring frame-level alignment of dialogue, expressions, and music. Extensive experiments demonstrate that CineAGI achieves 40\% improvement in overall consistency, 4.4\% gain in subject consistency, 5.4\% enhancement in aesthetic quality, and 28.7\% higher character consistency compared to baselines. Our work establishes a principled foundation for automated multi-scene video generation that preserves narrative coherence and character authenticity.
\end{abstract}

\section{Introduction}

\begin{figure*}[t]
\centering
\includegraphics[width=0.85\textwidth]{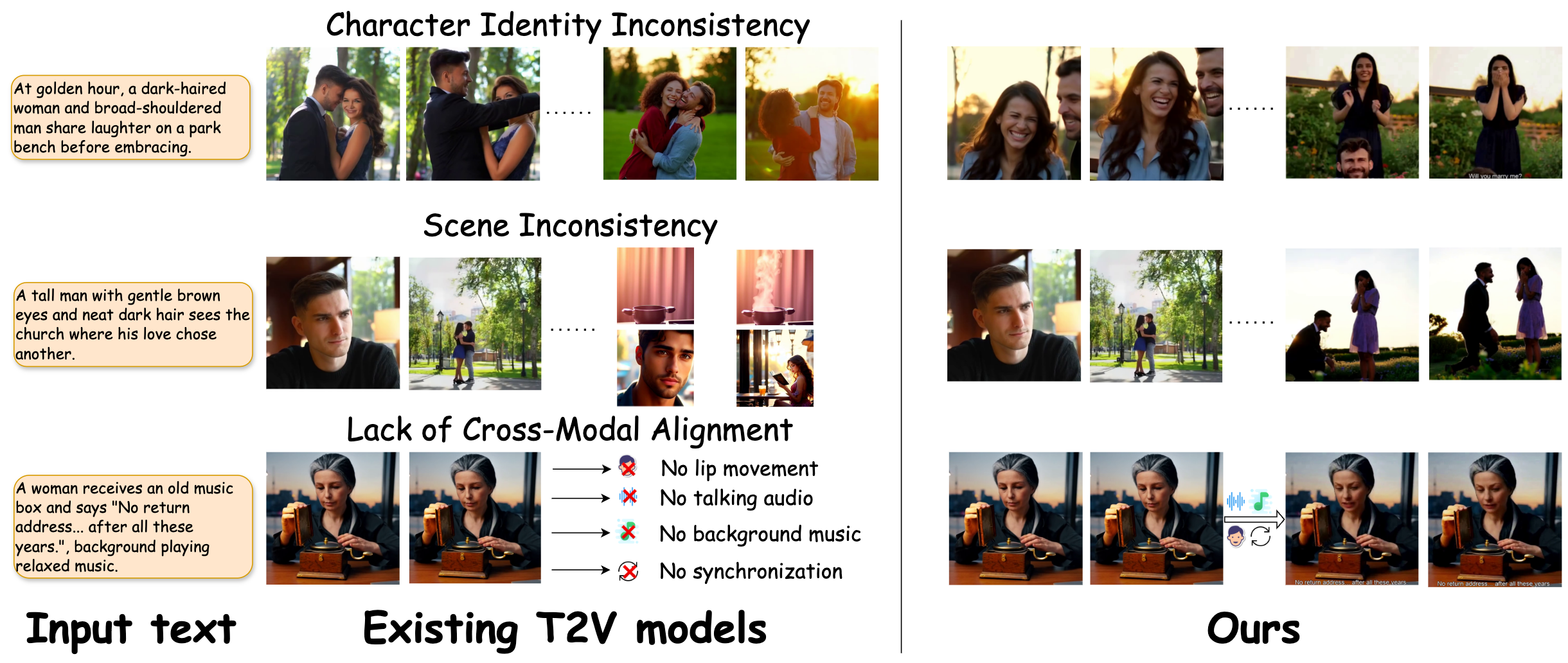}
\caption{\textbf{Key challenges in automated movie generation.} Existing text-to-video models face three fundamental limitations: (Top) \textit{Character identity inconsistency} across scenes and contexts; (Middle) \textit{Scene inconsistency} in visual styles and transitions; (Bottom) \textit{Cross-modal misalignment} between visual elements, dialogue, and audio. Our hierarchical framework addresses these challenges through specialized agent coordination and decoupled processing pipelines.}
\label{fig:teaser}
\end{figure*}

The automation of movie creation represents a fundamental challenge in generative AI, requiring the coordination of narrative structure, visual synthesis, and audio generation across extended temporal sequences. While recent advances in Large Language Models (LLMs)~\cite{ni2024autodirector,li2024anim} and diffusion-based video synthesis~\cite{rombach2022high,blattmann2023stable,peebles2023scalable} have shown impressive capabilities in individual content generation tasks, creating coherent long-form movies poses three critical challenges illustrated in Figure~\ref{fig:teaser}.

\textbf{Character Identity Preservation.} Quadratic computational complexity necessitates limited processing windows, causing loss of long-range identity information. Current approaches lack effective mechanisms to consistently preserve multi-scale identity features, from global attributes (facial structure, physique) to fine-grained details (expressions, lighting adaptation), across frames and scenes.

\textbf{Scene-Level Temporal Coherence.} Memory constraints prevent global scene consistency enforcement, and existing architectures lack explicit mechanisms for decomposing scene elements at different temporal frequencies, making it difficult to maintain compositional coherence over extended durations.

\textbf{Cross-Modal Synchronization.} Coordinating lip movements, emotional expressions, and musical timing at frame-level precision requires explicit synchronization mechanisms that existing end-to-end methods lack.

To address these challenges, we present CineAGI, a hierarchical movie creation framework that orchestrates multi-modal generation through principled decomposition and specialized agent coordination. Our framework introduces three key technical innovations:

\textbf{Multi-Agent Narrative Synthesis.} Specialized LLM agents (Character Designer, Script Writer, Storyteller, Composer, Quality Inspector) collaboratively generate cinematic blueprints. Unlike prior approaches that treat narrative elements independently, our coordinated agents maintain cross-modal consistency through structured information flow and validation mechanisms.

\textbf{Decoupled Character-Centric Pipeline.} A three-stage approach of text-guided segmentation (Grounded-SAM2), identity-preserving face integration (SimSwap), and talking face synthesis (Wav2Lip) enables independent multi-character processing while maintaining global consistency across diverse scene contexts.

\textbf{Hierarchical Audio-Visual Synchronization.} Explicit coordination at multiple levels, including frame-level lip alignment, emotion-aware music generation, and integrated scene assembly, addresses cross-modal alignment limitations in existing methods. This structured approach ensures synchronized audiovisual integration throughout the production.

Our main contributions are:
\begin{itemize}[leftmargin=*]
\item A novel multi-agent movie generation framework with specialized LLM agents for coordinated narrative planning, enabling fine-grained control over character consistency and temporal coherence that surpasses existing approaches.

\item A systematic character-centric pipeline that maintains multi-level identity preservation through decoupled processing, achieving robust character consistency across diverse contexts while preserving creative flexibility.

\item Comprehensive evaluation demonstrating 40\% improvement in overall consistency, 4.4\% gain in subject consistency, 5.4\% enhancement in aesthetic quality, and 28.7\% higher character consistency compared to state-of-the-art baselines.
\end{itemize}


\section{Related Work}

\subsection{Text-to-Video Generation}

Text-to-video synthesis has evolved rapidly through diffusion-based architectures. Recent systems such as Sora~\cite{brooks2024sora}, Lumiere~\cite{bartal2024lumiere}, and Stable Video Diffusion~\cite{blattmann2023stable} have significantly advanced generation duration and visual fidelity. CogVideoX~\cite{yang2024cogvideox} introduced expert transformers with 3D VAE for spatial-temporal compression, while VideoCrafter~\cite{chen2023videocrafter1} focused on temporal consistency through dedicated frame-level coherence modules. However, these methods face inherent limitations due to localized processing windows and lack of explicit narrative structure modeling.

Recent advances have addressed specific challenges in video generation. For character consistency, ID-Animator~\cite{he2024idanimator} and ConsisID~\cite{yuan2024consisid} introduced identity preservation techniques, while StoryDiffusion~\cite{zhou2024storydiffusion} and Animate Anyone~\cite{hu2024animateanyone} advanced long-range generation and character animation. For extended sequences, StreamingT2V~\cite{henschel2024streamingt2v} and FIFO-Diffusion~\cite{kim2024fifo} enabled multi-minute to theoretically infinite video generation. LLM integration has emerged as a powerful approach, with VideoDirectorGPT~\cite{lin2023videodirectorgpt} pioneering LLM-guided planning, VideoPoet~\cite{kondratyuk2024videopoet} demonstrating unified multimodal generation, and MovieGen~\cite{polyak2024moviegen} advancing joint video-audio synthesis. Despite these advances, maintaining multi-character consistency, cross-scene narrative coherence, and coordinated multi-modal production in long-form content remains challenging.

Our work differs fundamentally by introducing hierarchical decomposition through multi-agent orchestration that decouples character generation from scene synthesis, enabling robust character consistency while preserving narrative flexibility across extended sequences.

\subsection{Multi-Agent Systems}

Recent LLM advances have enabled sophisticated multi-agent systems for complex task automation. MetaGPT~\cite{hong2023metagpt} and AgentVerse~\cite{chen2023agentverse} demonstrated foundational principles for agent-based collaboration through specialized roles and structured coordination protocols.

In movie production, agent-based approaches have explored automated content creation. AutoDirector~\cite{ni2024autodirector} introduced multi-sensory composition through scheduled agents, while Anim-Director~\cite{li2024anim} focused on animation generation. AesopAgent~\cite{wang2024aesopagent} pioneered agent-driven evolutionary workflows combining RAG-based optimization with utility-layer execution, and StoryAgent~\cite{hu2024storyagent} introduced specialized agents mirroring professional production roles. However, these systems struggle with temporal consistency and multi-modal synchronization over extended sequences due to insufficient mechanisms for managing complex inter-dependencies.

Our hierarchical architecture advances beyond existing approaches through explicit coordination protocols inspired by professional pipelines, with specialized agents (Character Designer, Script Writer, Storyteller, Composer, Quality Inspector) ensuring both creative consistency and temporal coherence through structured information flow and validation mechanisms.
\begin{figure*}[t]
\centering
\includegraphics[width=0.9\linewidth]{framework.png}
\caption{\textbf{CineAGI framework overview.} Given a story concept, our framework proceeds through three modules: (1) \textit{Narrative Synthesis}: LLM agents collaboratively develop character profiles and production scripts ensuring narrative coherence; (2) \textit{Character Generation}: creates visual and audio assets with consistent character appearances and voice profiles; (3) \textit{Cinematographic Synthesis}: orchestrates video generation, character integration, and audio synchronization for complete movie production.}
\label{fig:framework}
\end{figure*}

\section{Method}

\subsection{Overview}

Our CineAGI framework introduces hierarchical multi-agent orchestration for automated movie creation, as illustrated in Figure~\ref{fig:framework}. Unlike end-to-end approaches that process entire sequences holistically, our hierarchical decomposition enables targeted optimization of individual components while maintaining global coherence through structured agent coordination.

The framework leverages complementary strengths of different generative models: LLMs for high-level creative planning and narrative structure, specialized models for consistent character generation, and targeted synthesis models for precise audiovisual integration. The \textit{Narrative Synthesis Module} serves as the creative foundation where LLM agents collaboratively develop rich, coherent story specifications. The \textit{Character Generation Module} bridges abstract descriptions and concrete assets through specialized portrait and voice synthesis. The \textit{Cinematographic Synthesis Module} coordinates these elements through decoupled processing streams, maintaining both local coherence and global consistency while precisely synchronizing visual, audio, and emotional elements.

\subsection{Narrative Synthesis Module}

Our narrative synthesis module introduces a hierarchical LLM agent framework that systematically decomposes movie creation into interconnected creative tasks. Unlike previous approaches~\cite{li2024anim,ni2024autodirector,zhuang2024vlogger} that treat narrative elements independently, our coordinated agents maintain cross-modal consistency through structured information flow and validation mechanisms.

\textbf{Character Designer} analyzes the input story to establish detailed character profiles, decomposing identities into hierarchical attributes covering appearance, personality, and behavioral patterns. These specifications serve as foundational references across all generation aspects, ensuring consistent character portrayal throughout the pipeline.

\textbf{Script Writer} develops shooting scripts centered around character profiles, maintaining strict alignment with Character Designer outputs. For each scene, it produces technical descriptions including visual composition, character positioning, camera movements, and detection keywords that guide downstream video synthesis.

\textbf{Storyteller} crafts narrative flows by analyzing character-scene relationships. It decomposes stories into coherent scenes, ensuring character actions align with established personalities while preserving dramatic arcs. It develops dialogue content with precise frame-level timing specifications to maintain consistent pacing and emotional progression.

\textbf{Composer} synthesizes character profiles, scene descriptions, and dialogue content to generate background music specifications. Rather than isolated scene-by-scene generation, it creates cohesive musical direction that enhances audiovisual synchronization and emotional resonance.

\textbf{Quality Inspector} ensures end-to-end consistency by validating interconnections between all agent outputs, preventing cascading errors and verifying relevance to the original input. It outputs structured JSON results that standardize inputs for subsequent processing stages. \textit{Detailed multi-agent coordination algorithms, protocol specifications, and convergence analysis are provided in Appendix A.}

\subsection{Character Generation Module}

The character generation module transforms abstract character profiles into concrete audiovisual assets. Operating on character profiles and dialogue scripts from the narrative stage, this module employs two specialized components:

\textbf{Portrait Artist} utilizes RealVisXL 3.0 to generate high-fidelity visual representations of each character. Taking detailed character profiles as input, it produces reference portraits capturing the character's appearance from multiple physical features. These portraits serve as primary references for maintaining visual consistency during face swapping in the cinematographic synthesis stage.

\textbf{Sound Generator} implements multi-identity voice synthesis via ChatTTS, combining character-specific voice profiles with dynamic emotional modulation. The framework processes comprehensive character specifications to generate voice assets that maintain speaker identity while supporting nuanced emotional expression. \textit{Technical specifications of the identity preservation mechanism and embedding consistency analysis are provided in Appendix B.}

\subsection{Cinematographic Synthesis Module}

The cinematographic synthesis module introduces a decoupled integration pipeline that systematically addresses character consistency and multi-modal synchronization:

\textbf{Scene Creator} employs HunyuanVideo-13B for text-driven scene generation. Unlike previous methods relying on character reference images, our framework encodes character specifications through rich textual descriptions, enabling flexible multi-character scene composition while providing a foundation for subsequent character-specific processing.

\textbf{Decoupled Character Integration} introduces a key technical innovation that transcends limitations of end-to-end single-character reference approaches. Our three-stage pipeline systematically decomposes multi-character scenes into individual segments while maintaining precise spatial-temporal relationships:

\textit{Character Segmentation} leverages Grounded-SAM2 for text-guided character isolation, processing scene-specific detection keywords to identify and track individual characters. This precise segmentation enables independent character processing while preserving contextual scene information.

\textit{Face Swapping} utilizes SimSwap~\cite{chen2020simswap} for identity-preserving face integration. By applying character-specific portrait references to segmented regions, this stage ensures consistent visual identity across diverse scene contexts.\textit{Comparative analysis of face swapping methods and identity preservation metrics are provided in Appendix C.}

\textit{Talking Face} synchronizes visual-audio modalities through Wav2Lip~\cite{prajwal2020lip}, using frame-level timing markers to coordinate character-specific lip movements with dialogue, ensuring natural facial dynamics while preserving both visual identity and audio synchronization.

\textbf{Music Virtuoso} leverages MusicGen to generate scene-specific background music based on musical directions from the Composer, synthesizing background music that aligns with each scene's emotional context while maintaining thematic coherence. The generated music adapts to scene durations and dramatic progression, complementing narrative structure without interfering with dialogue clarity.

\textbf{Cinematographer} executes final assembly through a multi-stage pipeline: integrating segmented characters back into original scene videos, overlaying character dialogue audio within specified frame ranges, adding corresponding subtitles, incorporating background music, and concatenating processed scenes according to narrative sequence. This systematic approach ensures synchronized audiovisual integration in the final output.

\section{Experiments}

\subsection{Experimental Setup}

\textbf{Evaluation Benchmark.} We construct a comprehensive benchmark comprising 100 diverse story prompts spanning five genres: romantic comedies, action sequences, dramatic narratives, family dramas, and suspense thrillers. Each prompt contains detailed character descriptions, plot elements, and emotional arcs to test different aspects of our system.

\textbf{Generation Settings.} For fair comparison, we sample videos at 24 FPS with 5.375-second duration (129 frames per scene) at 512×512 resolution. Baseline models are evaluated using two complementary strategies: (1) generating multiple scenes with different random seeds until matching our total video length; (2) using scene descriptions from our Script Writer to generate equivalent-length videos per scene. Quantitative metrics are computed as average scores across both strategies, providing comprehensive evaluation accounting for both generation approaches.

\textbf{Evaluation Metrics.} We employ the standardized VBench framework to assess multiple aspects: Overall Consistency (OC) through ViCLIP evaluates text-to-video alignment and story flow continuity; Subject Consistency (SC) assesses maintenance of character identities across scenes; Aesthetic Quality (AQ) evaluates visual fidelity and artistic composition; Motion Smoothness (MS) measures fluidity of character movements and scene transitions.

\textbf{Baselines.} We compare against CogVideoX~\cite{yang2024cogvideox}, VideoCrafter2~\cite{chen2024videocrafter2}, and Hunyuan~\cite{li2024hunyuan}.

\textbf{Computational Cost.} Our full pipeline requires approximately 11.3 minutes per scene (5.375s) on a single NVIDIA A100 GPU; a detailed per-component breakdown is provided in Appendix~E.

\begin{figure*}[t]
\centering
\includegraphics[width=0.85\linewidth]{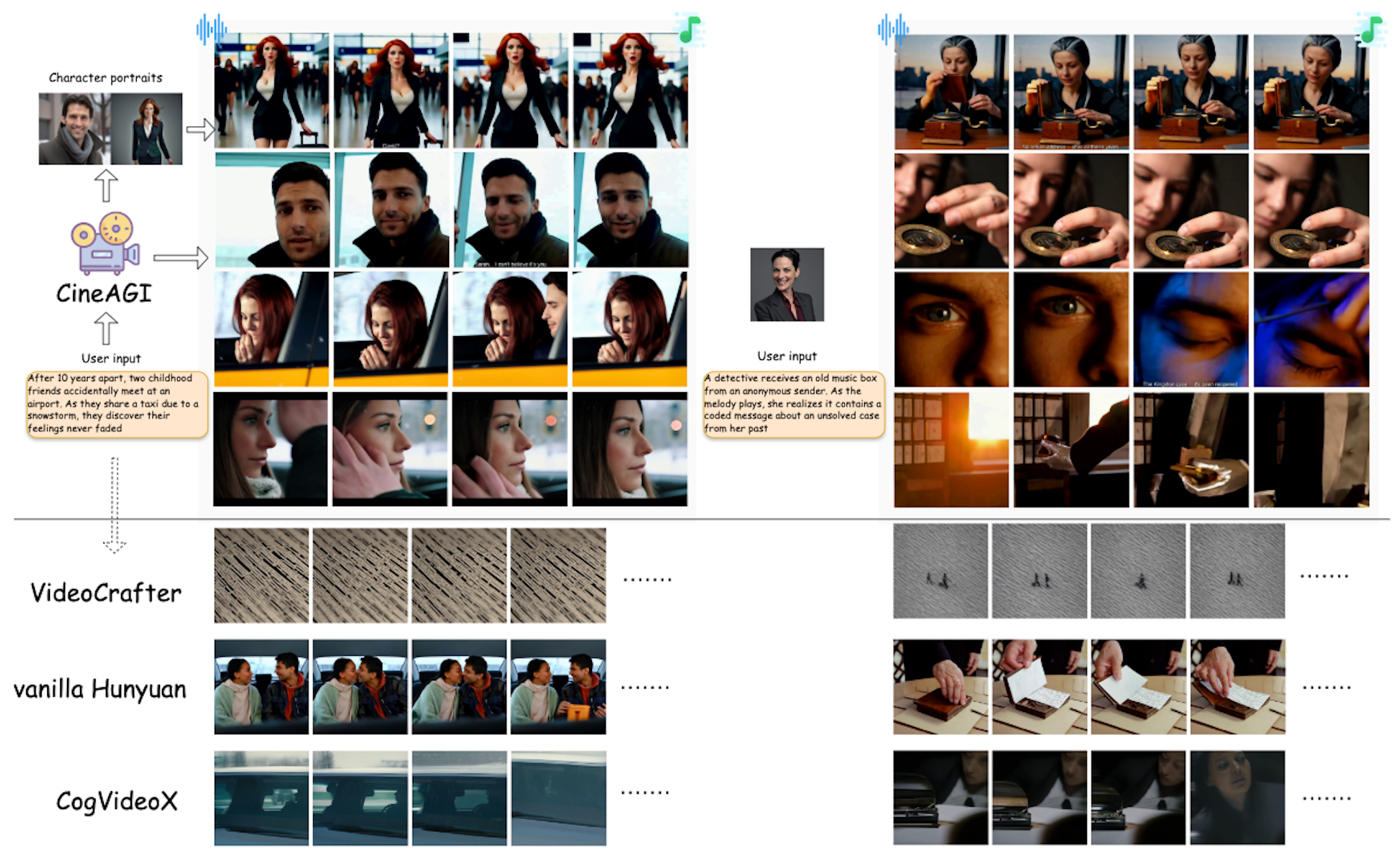}
\caption{\textbf{Qualitative comparisons with state-of-the-art methods.} Our approach generates more coherent narratives with consistent character portrayals across varied scenes while maintaining superior visual quality and natural character movements.}
\label{fig:qualitative}
\end{figure*}

\subsection{Quantitative Evaluation}

Table~\ref{tab:quantitative} presents quantitative results. CineAGI achieves the highest Overall Consistency (0.259, 40\% improvement over Hunyuan), Subject Consistency (0.949, 4.4\% improvement), Aesthetic Quality (0.600, 5.4\% improvement), and Motion Smoothness (0.987, 1.1\% improvement), demonstrating superior narrative coherence and character consistency. \textit{Per-genre performance breakdown, failure case analysis, and computational efficiency comparison are provided in Appendix E.}

\begin{table}[t]
\centering
\caption{\textbf{Quantitative comparison.} Our method achieves superior performance across all metrics. OC: Overall Consistency, SC: Subject Consistency, AQ: Aesthetic Quality, MS: Motion Smoothness.}
\label{tab:quantitative}
\begin{tabular}{l|cccc}
\toprule
\textbf{Method} & \textbf{OC}$\uparrow$ & \textbf{SC}$\uparrow$ & \textbf{AQ}$\uparrow$ & \textbf{MS}$\uparrow$ \\
\midrule
CogVideoX        & 0.096 & 0.823 & 0.379 & 0.960 \\
VideoCrafter2    & 0.076 & 0.885 & 0.364 & 0.920 \\
Hunyuan          & 0.185 & 0.909 & 0.569 & 0.976 \\
\midrule
\textbf{CineAGI} & \textbf{0.259}$^{**}$ & \textbf{0.949}$^{**}$ & \textbf{0.600}$^{*}$ & \textbf{0.987}$^{*}$  \\
\textbf{Improvement} & \textbf{+40.0\%} & \textbf{+4.4\%} & \textbf{+5.4\%} & \textbf{+1.1\%} \\
\bottomrule
\multicolumn{5}{l}{\footnotesize $^{*}p < 0.05$, $^{**}p < 0.01$ (paired t-test over 100 prompts vs.\ best baseline).}
\end{tabular}
\end{table}

\subsection{Human Evaluation}

We conducted a user study with 20 participants (10 with professional multimedia experience) rating videos on a 5-point Likert scale across five dimensions.

Table~\ref{tab:user_study} shows superior performance across all dimensions. CineAGI achieves highest scores in Visual Quality (3.83), Narrative Coherence (3.57), and Character Consistency (3.14, 28.7\% improvement). Audio Coherence (3.26) demonstrates effective multi-modal integration unique to our approach.

\begin{table}[t]
\centering
\caption{\textbf{Human evaluation results.} Scores rated from 1-5 (higher is better). VQ: Visual Quality, NC: Narrative Coherence, CC: Character Consistency, AC: Audio Coherence, OQ: Overall Quality. (-) indicates unsupported feature.}
\label{tab:user_study}
\begin{tabular}{l|ccccc}
\toprule
\textbf{Method} & \textbf{VQ}$\uparrow$ & \textbf{NC}$\uparrow$ & \textbf{CC}$\uparrow$ & \textbf{AC}$\uparrow$ & \textbf{OQ}$\uparrow$ \\
\midrule
CogVideoX        & 3.16 & 2.52 & 2.21 & - & 2.63 \\
VideoCrafter2    & 2.75 & 2.26 & 1.98 & - & 2.45 \\
Hunyuan          & 3.52 & 2.91 & 2.44 & - & 2.88 \\
\midrule
\textbf{CineAGI}    & \textbf{3.83} & \textbf{3.57} & \textbf{3.14} & \textbf{3.26} & \textbf{3.37} \\
\textbf{Improvement} & \textbf{+8.8\%} & \textbf{+22.7\%} & \textbf{+28.7\%} & - & \textbf{+17.0\%} \\
\bottomrule
\end{tabular}
\end{table}

\subsection{Ablation Study}

\begin{figure}[t]
\centering
\includegraphics[width=0.9\linewidth]{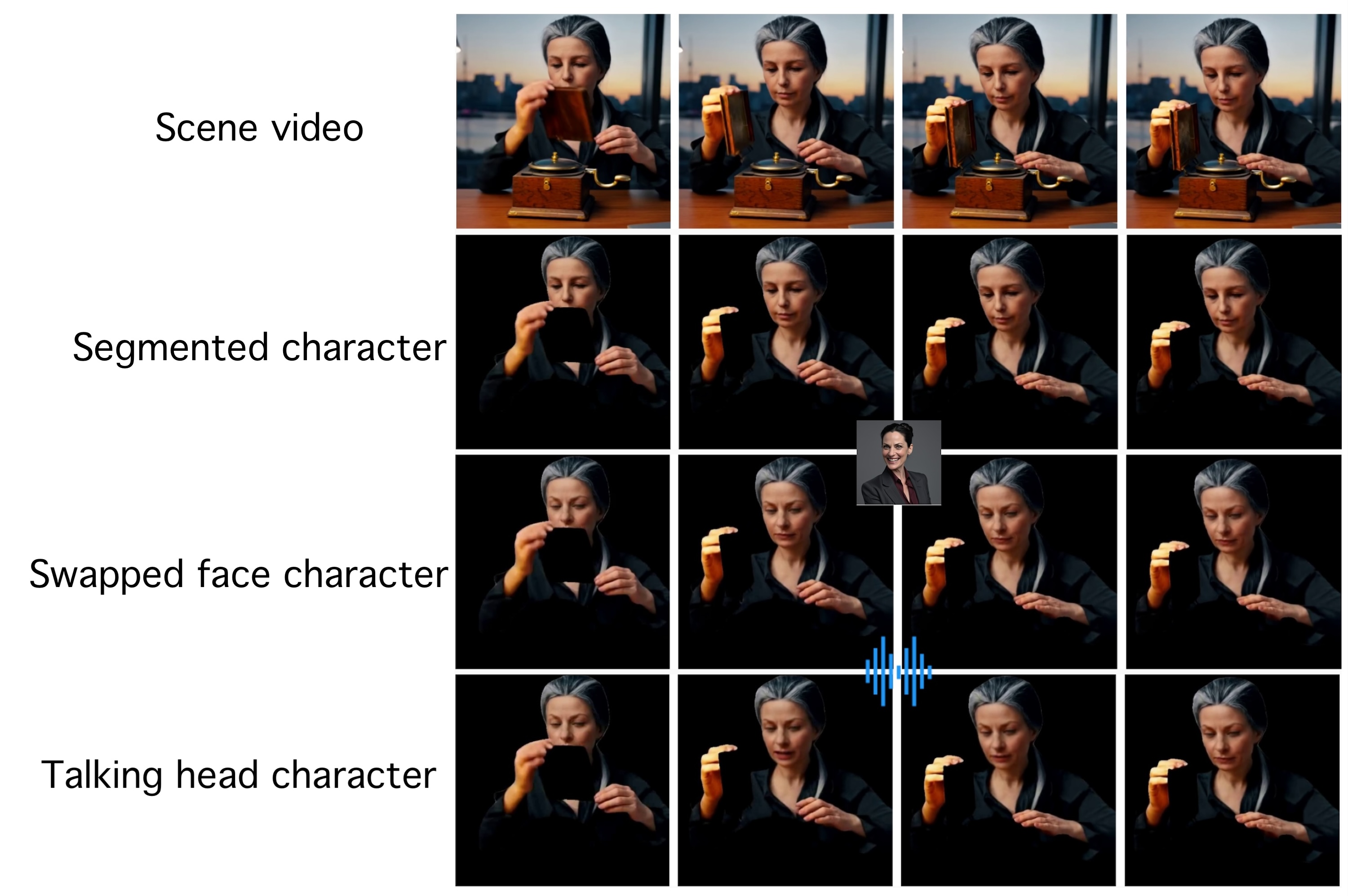}
\caption{\textbf{Decoupled Character Integration (DCI) pipeline visualization.} From top to bottom: original scene video, segmented character regions, face-swapped results with reference portrait, and final talking head output with audio synchronization. This systematic decomposition enables precise character control while maintaining scene context.}
\label{fig:ablation}
\end{figure}

We conduct comprehensive ablation studies to validate our design choices (Table~\ref{tab:ablation}, Figure~\ref{fig:ablation}). We examine three key components:

Table~\ref{tab:ablation} validates our design choices. Without Narrative Synthesis Module, Overall Consistency drops to 0.232 and Subject Consistency to 0.924. Without Quality Inspector, performance moderately decreases (OC: 0.245, SC: 0.938). Without Decoupled Character Integration, Aesthetic Quality decreases to 0.583 and Motion Smoothness to 0.971, confirming each component's essential role. \textit{Individual agent contribution analysis, alternative architecture comparisons, and sensitivity analysis for key hyperparameters are provided in Appendix F.}




\begin{table}[t]
\centering
\caption{\textbf{Ablation study results.} NSM: Narrative Synthesis Module, QI: Quality Inspector, DCI: Decoupled Character Integration.}
\label{tab:ablation}
\begin{tabular}{l|cccc}
\toprule
\textbf{Variant} & \textbf{OC}$\uparrow$ & \textbf{SC}$\uparrow$ & \textbf{AQ}$\uparrow$ & \textbf{MS}$\uparrow$ \\
\midrule
w/o NSM     & 0.232 & 0.924 & 0.575 & 0.974 \\
w/o QI      & 0.245 & 0.938 & 0.570 & 0.982 \\
w/o DCI     & 0.206 & 0.911 & 0.583 & 0.971 \\
\midrule
\textbf{Full CineAGI} & \textbf{0.259} & \textbf{0.949} & \textbf{0.600} & \textbf{0.987}  \\
\bottomrule
\end{tabular}
\end{table}

\subsection{Qualitative Results}

Figure~\ref{fig:qualitative} demonstrates CineAGI's effectiveness in producing coherent movies with consistent character portrayals and appropriate cinematography.


\section{Conclusion}

We present CineAGI, a hierarchical framework for automated movie creation that coordinates multi-agent narrative synthesis, decoupled character-centric processing, and audio-visual synchronization. Extensive experiments validate significant gains across both automated metrics and human evaluation, establishing a modular foundation for AI-driven filmmaking.

\bibliographystyle{IEEEbib}
\bibliography{icme2026references}


\end{document}